\documentclass[%
 aip,
 amsmath,amssymb,
 reprint,%
]{revtex4-1}

\usepackage{graphicx}% Include figure files
\usepackage{dcolumn}% Align table columns on decimal point
\usepackage{bm}% bold math
\usepackage[utf8]{inputenc}
\usepackage[T1]{fontenc}
\usepackage{mathptmx}

\def\PRL{{Phys. Rev. Lett.} }
\def\PRA{{Phys. Rev.} A }

\newcommand{\be}{\begin{equation}}
\newcommand{\ee}{\end{equation}}
\newcommand{\bea}{\vspace{0.25cm}\begin{eqnarray}}
\newcommand{\eea}{\end{eqnarray}}

\begin{document}

\preprint{AIP/123-QED}

\title[]{Experimental Quantum Enhanced Optical Interferometry}
% Force line breaks with \\

\author{M.Genovese}

\affiliation{
Istituto Nazionale di Ricerca Metrologica, Strada delle Cacce 91, 10135, Torino, Italy and  INFN, sezione di Torino, via P. Giuria 1, 10125 Torino, Italy
}

\date{\today}% It is always \today, today,

\begin{abstract}
Optical quantum interferometry represents  the oldest  example of quantum metrology and it is at the source of quantum technologies.
The original squeezed state scheme is now a significant element of the last version of gravitational wave detectors and various additional uses have been proposed.
 Further quantum enhanced schemes, from SU(1,1) interferometer to twin beam correlation interferometry, have also reached the stage of proof of principle experiments enlarging the field of experimental quantum interferometry and paving the way to several further applications  ranging from Planck scale signals search to small effects detection.
In this review paper I  introduce these experimental achievements, describing their schemes, advantages, applications and possible further developments.
\end{abstract}

\maketitle

\section{\label{sec:level1}Introduction}

In the last two decades a second quantum revolution based on exploiting peculiar properties of single quantum states prompted the emerging of quantum technologies, ranging from quantum information to quantum metrology \cite{qt}.
Substantial progresses have been made in the development of quantum-enhanced measurement in recent years: Quantum metrology  is the new discipline addressed to overcome the limits of classical measurements, in particular shot noise \cite{sn}, by exploiting specific properties, and in particular peculiar correlations, of quantum systems, as entanglement \cite{1}, discord \cite{2} or squeezing \cite{3}.
As other quantum technologies (e.g. quantum information, quantum computation, quantum communication, …) it had an exponential growth in the last 20 years, leading to the realization of several interesting proof of principle experiments, mostly in quantum optics, and approaching now practical applications ranging from biology to remote detection \cite{QSRev,ImRev,Giov04,Ol18,cap,bio}. As  demonstrated potentialities of this discipline one can mention, without any pretension to be exhaustive, the possibility of enhancing the performances of interferometers \cite{Int,mc,IRev,IRev1,IRev2,IRev3,Iv,Bot00,Hu}, of phase estimation \cite{PE}, of object detection or testing \cite{qillu,qread}, of superresolution \cite{lit,lit2}, of spectroscopy  \cite{SPE} and of beating shot noise in imaging \cite{SSNIm,lS,ImRev} or absorption measurements \cite{SSNAb,abs}.

In particular, the application of quantum enhanced schemes to optical  interferometry, a technique of huge widespread application ranging from basic science \cite{ligo,virgo} to computation \cite{comp},  represents  the oldest  example of quantum metrology and it is at the very source of quantum technologies. Nowadays, the advantage of using squeezed light in interferometers has found application in enhancing the sensitivity of the upgraded version of gravitational wave experiments as GEO 600 \cite{geo1,geo2}, LIGO \cite{ligo}  and VIRGO \cite{virgo}, while new ideas are reaching a  proof of principle stage. In this review paper I will introduce the main elements of Experimental Quantum Enhanced Optical Interferometry, touching the following arguments:

- General introduction to quantum enhanced interferometry

 - Squeezed light enhanced interferometers

- SU(1,1) interferometers

- Quantum enhanced correlation interferometers

- New ideas and perspectives

- Conclusions

Further applications of quantum interferometry addressing quantum imaging, absorption measurement etc. are beyond the puropose of this paper and can be found, for instance, in Ref. \cite{ImRev,nl}

\section{\label{sec:level1}General introduction to quantum enhanced interferometry}

Nowadays interferometers represent probably  the most sensitive instruments we dispose of.
Nevertheless, several kinds of noise must be coped for reaching the highest sensitivity\cite{rev}. Just for mentioning the ones considered by LIGO collaboration \cite{ligo2}:  Thermal noise, Seismic noise, Newtonian noise, Laser frequency noise, Laser intensity noise, Auxiliary length control noise, Actuator noise, Alignment control noise, Beam jitter noise, Scattered light noise, Residual gas noise, Photodetector dark noise,...

Anyway even when all these sources of noise are tamed, intrinsic limits remain. In particular the classical interferometer sensitivity scales as $1 / \sqrt{N}$ where N is number of photons (i.e. scales with the square root of the inputed light intensity), the so called shot noise.

Surprisingly enough this limit can eventually be beaten by exploiting  peculiar quantum properties of light \cite{illu}, i.e. by inputing light states with specific properties of quantum systems (as  squeezing or entanglement) \cite{prep,fabre}. In this last case the phase sensitivity $\Delta \theta$ can eventually reach the so called Heisenberg limit, i.e. a $1/N$ scaling. A limit generically  imposed by the laws of quantum mechanics, namely, the generalized Heisenberg uncertainty $\Delta N \Delta\theta \geq 1$ \cite{h1,h2,h3,h4,h5,h6,h7,h8,h9,h10,h11}. It is worth noticing that even this limit can eventually be beaten in presence of prior information \cite{sh1,sh2} or should be refined in specific situations, indeed Hofmann \cite{Hof} suggested to consider the limit
\be
\triangle \theta^2 \geq 1/ \langle n^2 \rangle
\ee
where one considers the average of the squared photon number $\langle n^2 \rangle$. This allows for a better sensitivity in the case of large photon number fluctuations, $\triangle n^2 = \langle n^2 \rangle -\langle n \rangle^2 > 0$. However, the effective possibility of using this advantage is still under discussion \cite{VL}.

The possibility of beating classical limits in interferometry represents one of the most intriguing chances offered by quantum metrology, that now is overpassing the stage of theoretical proposals \cite{demo}, reaching experimental applications (in some case beyond or proof of principle demonstrations), whose a few examples will be discussed in the following.

\section{\label{sec:level1}Squeezed light enhanced interferometers}
As mentioned the first proposal of quantum enhanced measurement was  the idea of Caves \cite{Int} concerning the use of squeezed states for improving the performances of interferometers under the shot noise.
In real situations eventually other technical noises can be more important than shot noise, but, as we will see, this proposal finally found an important application in gravitational wave detectors.

The idea is relatively simple. In a "traditional" interferometer a laser beam (well approximated by a coherent state) enters a port of the system, propagates inside the interferometer split in two (or eventually more) components that at end are recombined and interfere according to the relative acquired phase $\theta $.

\begin{figure}
\includegraphics[height=6cm]{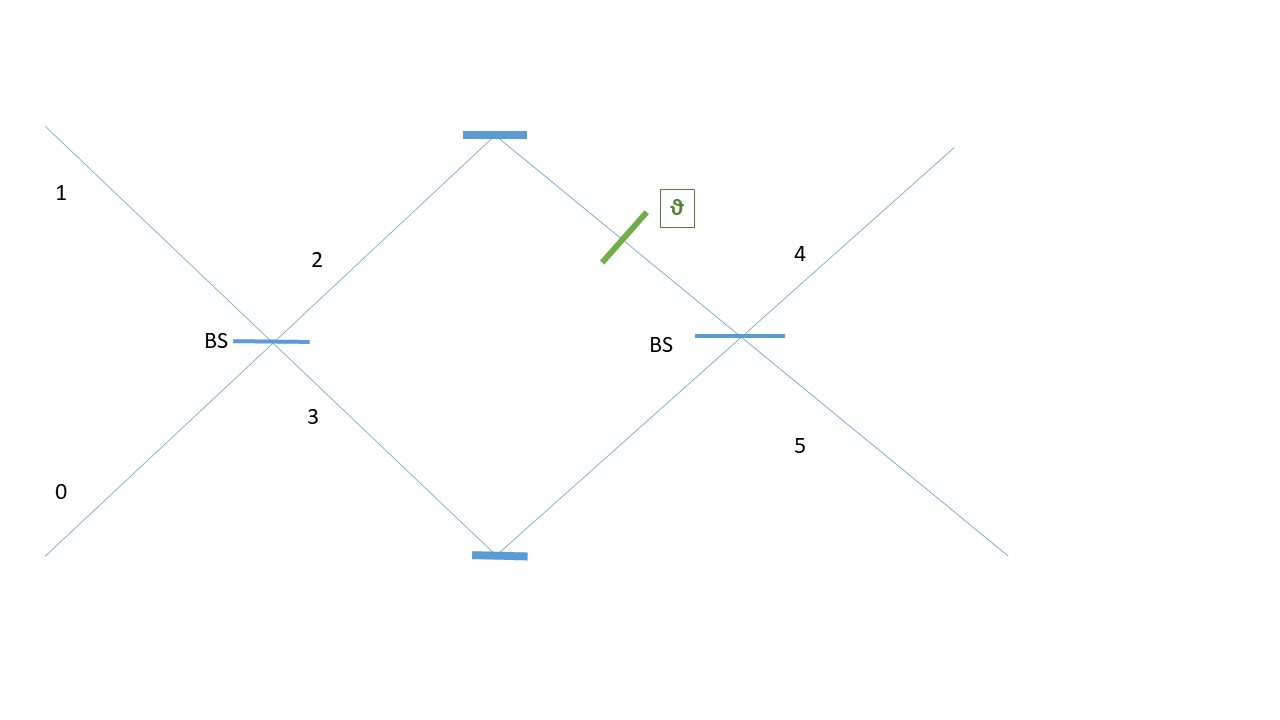}% Here is how to import EPS art
\caption{\label{mz} Schematics of a Mach Zehnder interferometer.}
\end{figure}

Let us consider for example a Mach-Zehnder interferometer (but the same considerations apply mutate mutandis to any other interferometer), see fig.\ref{mz}.
Let call 0 and 1 the two modes entering the initial beam splitter. After propagation in the two arms, acquiring a relative phase $\theta $, they recombine on a second beam splitter, whose outputs we denote with modes 4 and 5.
The photon destruction operator corresponding to mode 5 can thus be written in terms of the operators relative to the input modes 0,1 as:

\begin{equation}
a_5 = i e^{ i \theta /2}  \cdot ( a_1 \sin{ ( \theta /2)} + a_0 \cos {( \theta /2)} )
\end{equation}

Thus, if the mode 1 is a coherent state $ | \alpha \rangle = D(\alpha) | vacuum \rangle$ (much more intense than the state inputing mode 0), where $  D(\alpha) = \exp ( \alpha  a - \alpha ^* a^\dag )$, the output photon number in mode 5, $\langle n_5 \rangle = \langle a_5^{\dag} a_5 \rangle$ is:

\begin{equation}
\langle n_5 \rangle = \sin ( \theta /2)^2 | \alpha |^2 + \sin ( \theta /2) \cos ( \theta /2) | \alpha | \langle a_0 e^{-i \theta} + a_0^\dag e^{i \theta} \rangle
\end{equation}

$ X_{\theta} = 1/\sqrt{2} ( a_0 e^{-i \theta} + a_0^\dag e^{i \theta})$ being the quadrature variable (in the following $X_0 = X$).

If, on the same line, we evaluate the variance of $n_5$ we arrive to
\be
\langle \Delta n_5 ^2 \rangle = (\sin ( \theta /2) \cos ( \theta /2) | \alpha | )^2 \langle \Delta X^2 \rangle
\ee
If the port 0 is unused, i.e. the input is the vacuum state, $| vacuum \rangle$, then $\langle \Delta X^2 \rangle = 1/4$.
This represents the shot noise.

Caves demonstrated \cite{Int} that  inputing on the port 0 a squeezed state $ D(\alpha) S(z)  | vacuum \rangle $, where $S(z) = \exp [ 1/2 (a^2 z^* - (a^\dag)^2 z)]$, $z = r e^{(i \phi)}$, e.g. the squeezed vacuum $S(z)  | vacuum \rangle $, then
\be
\langle \Delta X^2 \rangle =  e^{-(2 r)}/4 \, \,\, ,
\ee
i.e. the noise is reduced respect to the shot noise (a detailed mathematical analysis of shot noise, and in general noise in interferometers, can be found in Ref.\cite{rev}).

 This result represented a fundamental progress, showing as the use of quantum resources, in the case the squeezing, can lead to an improvement respect to classical limits, in this case the shot noise.

\begin{figure}
\includegraphics[height=6.5cm]{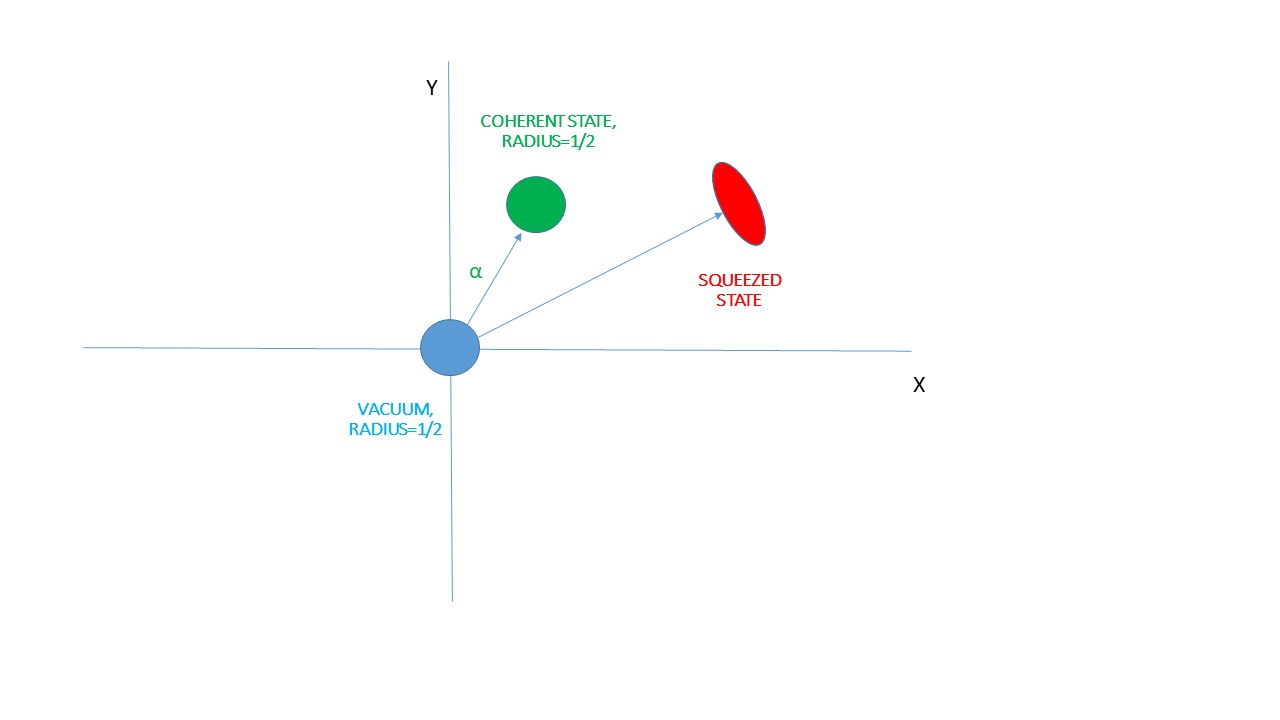}
\caption{\label{COH-SQ}Uncertainty circle in complex-amplitude state for vacuum, coherent state $| \alpha \rangle$ and squeezed state.}
\end{figure}

This upshot is related to the specific quantum properties of squeezed states. When considering quadrature variables, $X= 1/\sqrt{2} (a + a^\dag)$   and $Y  = 1/(\sqrt{2}i) (a - a^\dag)$, for the vacuum $\Delta X = \Delta Y = 1/2 $, the same is for coherent states, which in the complex-amplitude plane are just a translated vacuum (see fig.\ref{COH-SQ}). On the other hand, for squeezed states the variance in one quadrature is "squeezed", e.g. $\Delta X = 1/2 e^{-r}$, while for the other increases, $\Delta Y =  1/2 e^{r}$, keeping the product fixed to 1/4. By combining a squeezed and a coherent state on a beam splitter antibunched light can be produced, where eventually there are increased amplitude fluctuations and better defined phases \cite{Int}.  Incidentally, by combining two of them  a photon number entangled state is obtained (a two mode squeezed state) \cite{kim01}.

 For several years this proposal remained of theoretical interest, or had proof of principle experiments \cite{s1,s2}. Nonetheless, in the last years the situation strongly changed. After that in Ref.\cite{mc} was experimentally demonstrated a  2.3 dB  improvement in a power recycled Michelson with squeezed light injected into the dark port, this method found a large interest  with the development of extremely sensitive interferometers for gravitational wave search.
 Following the use in Geo 600 \cite{geo1,geo2}, it found also application in LIGO \cite{ligo}, where it now represents a  significant component \cite{ligo2}.

 Respect to previous versions, now in LIGO the squeezed vacuum source (an optical parametric oscillator, OPA) is placed inside the vacuum envelope on a separate suspended platform, reducing in this way the squeezing ellipse phase noise
and backscattered light noise. Furthermore, the squeezer has been fully integrated into the automated lock acquisition sequence. Thanks to this improvements, now above 50 Hz the interferometer sensitivity is increased
by 2.0 dB and 2.7 dB at LIGO Hanford Observatory and LIGO Livingston Observatory, respectively. Thus, while increasing  interferometer sensitivity by increasing the input power to the interferometer, which grows by enhancing the gravitational-wave signal, is limited by radiation pressure inducing instabilities and absorption of the test masses,  injecting squeezed vacuum improves the signal-to-noise ratio by decreasing the interferometer noise. In this sense 3 dB of squeezing is equivalent to doubling the intracavity power to 450kW. For instance, this provides a 12\% and 14\% increase in binary neutron star in spiral range at each respective site \cite{ligo2}. Similar improvements have also been obtained into the other big gravitational wave detector, Virgo \cite{virgo}.
Nonetheless, below 50 Hz, injecting frequency-independent squeezed vacuum increases the quantum radiation pressure noise This effect in LIGO  limits at the moment a further increasing of the current squeezing level. Frequency dependent squeezing could alleviate this problem.

Beyond these technical limits (e.g. contribution to radiation pressure noise), also losses are a significant problem \cite{losses}, since they reduce the squeezing\cite{leu,x}.  The conundrum of keeping a strong squeezing coping losses without increasing the squeezing level, can eventually be solved by amplifying the squeezing quadrature where the information is encoded \cite{a1}. A proof of principle of this scheme has been recently realised with a Mach Zehnder like interferometer \cite{a2}, demonstrating a 6dB sub shot noise sensitivity even for $50\%$ detection efficiency.

Coming back to the tradeoff between radiation pressure noise and shot noise reduction, some idea for improving both by realising frequency dependent squeezing has recently emerged. The point is that quantum fluctuations in the amplitude quadrature of the light generate a fluctuating radiation pressure on mirrors, that in turn couples amplitude and phase quadratures \cite{eprSch}
\be
a^{phase}_{out} = a^{phase}_{in} - K a^{ampl}_{in} + signal
\ee
where the Kimble factor $K$ depends on the frequency (beyond the line width of detector, the light power, the mass of mirror).

The radiation pressure noise (dominating at low frequency) can be reduced by squeezing the amplitude quadrature, the shot noise (dominating the high frequency sensitivity) demands for squeezing the phase quadrature. In Ref. \cite{eprSch1} it was suggested a method for realising this situation, based on generating EPR entangled beams \cite{prep} through a detuned OPA. The method takes advantage of the entanglement between fields around the half of the pump frequency: measuring on the correlated modes allows reducing the uncertainty on the other (conditional squeezing), while being idler and signal fields separated by tens of MHz, it will not mix with the strong carrier to produce radiation pressure on  mirrors. This scheme has been recently demonstrated in a proof of principle experiment \cite{eprSch}.

Further semiclassical methods for improving gravitational wave detectors, in particular for getting rid of back-action fluctuations, can be found in \cite{rev}, such as speed meter interferometers \cite{s1,s2,s3,s4,s5,s6,s7,s8,s9,s10,s11} ( performing quantum non demolition measurements of mirror velocity), coupling to another (quantum enhanced) interferometer \cite{s12,s13} or white-light cavity schemes (i.e. broadening the bandwith) \cite{s14}.

Before concluding this section, it is worth mentioning that a further theoretical progress was the demonstration of Ref.\cite{smerzi} that the choice of the average relative number of photons as a phase estimator is not
optimal (albeit the simplest one and "feasible" with current technology), since further information about the true value of the phase shift is contained in the quantum fluctuations of the number of particles measured at the output ports. In summary, the combination of coherent and squeezed light at input beam splitter creates an entangled state that a suited measure could further exploit. Theoretical considerations, based on Cram\'er-Rao bound,  lead to a bound (p being the number of measurements and $F(\theta)$ the Fisher information)

\be
\Delta \theta =  {1 \over \sqrt{p F(\theta)}} = {1 \over \sqrt{p (|\alpha|^2 e^{2r}+ \sinh^2(r))}} \,\,
\ee

that reaches the Heisenberg scaling
\be
\Delta \theta =  {1 \over \sqrt{p} N}
\ee

when $|\alpha|^2 \simeq \sinh^2(r) \simeq N/2$ and $p, N << 1$.

Finally, the possibility of improving interferometry in presence of losses by exploiting a phase-sensitive amplifier on the outputs was discussed and experimentally demonstrated for a coherent state input in \cite{Spa}, while  further a few theoretical ideas, that could lead to improvements in a more far future, emerged, concerning the use of intelligent states \cite{i1,i2,i3},  extended squeezed states  \cite{th1} or other methods  \cite{th2,th3,th4,th5}.

\section{\label{sec:level1}SU(1,1) interferometers}

Another interesting class  of quantum enhanced interferometers is represented by nonlinear interferometers \cite{nl}, and in particular the SU(1,1) one \cite{ou}, whose name derives from the different transformation performed respect to the SU(2) -type interaction realised by a beam splitter in a traditional interferometer. Here the beam splitters are substituted by parametric amplifiers (PA), where the two outputs (usually dubbed idler and signal) of the first PA are recombined in the second PA. This scheme was then demonstrated to be able to reach the Heisenberg scaling 1/N (in terms of photon number N), Ref.\cite{Yurke} .

Indeed, when defining the generators of SU(2) in terms of input mode annihilation operators
\bea
J_1=(a_0^\dag a_1 + a_1^\dag a_0)/2 \\
J_2 =- i (a_0^\dag a_1 - a_1^\dag a_0)/2 \\
J_3 = (a_0^\dag a_0 - a_1^\dag a_1)/2
\eea
the beam splitter with transmissivity $\cos(\alpha /2)$  corresponds to the related group transformation $U=e^{-i \alpha J_1}$.

On the other hand, by defining the SU(1,1) algebra generators
\bea
K_1=(a_0^\dag a_1^\dag + a_1 a_0)/2 \\
K_2 =- i (a_0^\dag a_1^\dag - a_1 a_0)/2 \\
K_3 = (a_0^\dag a_0 + a_1 a_1^\dag)/2
\eea
the PA is described by the relative group element $U=e^{-2i r K_2}$.

By considering the sequence of transformations describing a traditional interferometer (Mach-Zehnder, Fabry-Perot or whatelse) one obtains the usual $\backsim 1/\sqrt{N}$ sensitivity \cite{Yurke}; while,
as the squeezed state interferometers, also SU(1,1) interferometers can achieve a phase sensitivity of 1/N, but in this case with only vacuum fluctuations entering the input ports and coherent states pumping the active devices.

The original proposal was based on spontaneous emission (i.e. the parametric amplifier had the input modes as vacuum states): this is not very effective due to the limitation in the intensity (i.e. in N) of the output modes.
Thus, experimental realisations follow a scheme where coherent states are injected \cite{c1,c2}.

The experimental set-up, and the possible eventual applications, depend on the medium chosen as parametric amplifier. At the moment three different configurations have been realized: exploiting four-wave mixing \cite{fwm1,fwm2,fwm3,fwm4,fwm5}, a bulk crystal \cite{b} or non-linear optical fibers \cite{nf1,nf2,nf3}.

The general idea of a SU(1,1,) interferometer is, as mentioned, to substitute the beam splitters with parametric amplifiers, see fig\ref{su}, whose Hamiltonian is (in parametric approximation where the pump field is approximated to a classical field)
\be
H_{PA} = i ( g a_1^\dag a_2^\dag - g^* a_1  a_2 )
\ee
\begin{figure}
\includegraphics[height=6cm]{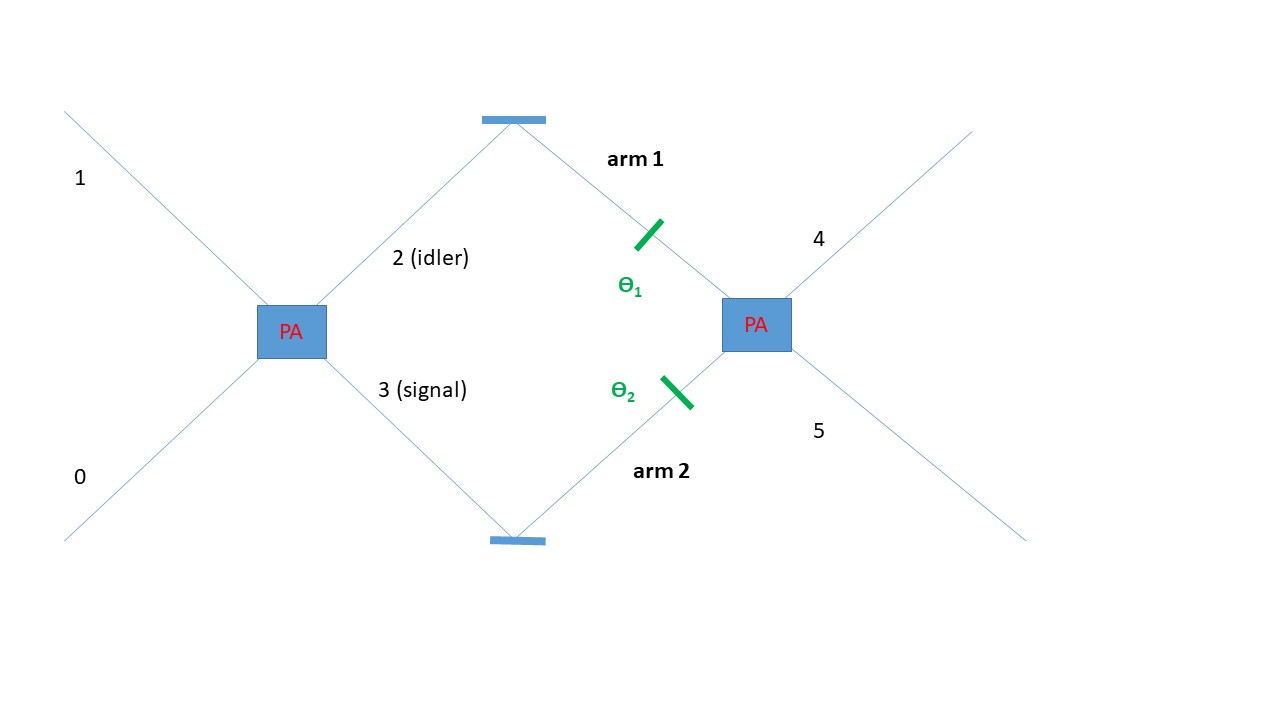}% Here is how to import EPS art
\caption{\label{psd} Schematics of a SU(1,1) interferomter.}
\end{figure}
Introducing the gain $G_j = \cosh (k g)$, k being a constant, with $j=1,2$ denoting the first or second PA, when a coherent state  $| \alpha \rangle$ is inputed in mode 1 (while the other input mode is in the vacuum) the output intensities are:
\begin{align}
I_4 = & |\alpha|^2 ( G_1^2 G_2^2 + (G_1^2-1) (G_2^2-1) + \nonumber \\ & 2 G_1 G_2  \sqrt{G_1^2-1} \sqrt{G_2^2-1} \cos {(\theta_1 + \theta_2)}
\end{align}
\begin{align}
I_5 = & |\alpha|^2 ( G_1^2 \cdot (G_1^2-1)+  G_2^2 \cdot (G_2^2-1) + \nonumber \\& 2 G_1 G_2  \sqrt{G_1^2-1} \sqrt{G_2^2-1} \cos {(\theta_1 + \theta_2)}
\label{su}
\end{align}
where $\theta_1$ and $ \theta_2$ are the phases acquired on the arm 1 and 2 of the interferometer, respectively.

One interesting property that differentiates a SU(1,1) interferometer from a usual Mach-Zehnder is, as one immediately evinces form Eq.\ref{su}, that the interference fringes depends on the sum of  $\theta_1$ and $ \theta_2$ (instead of depending on the difference). More interesting  the fringe size depends quadratically (thanks to amplification) on the phase sensing field strength to be compared to a linear dependence of the linear interferometer.
Finally, another significant difference is that for the SU(1,1) interferometer making the difference between the two outputs the phase dependence is cancelled.

However, the most interesting property of a   SU(1,1) interferometer  is that the noise is not amplified as much as the signal, increasing the signal to noise ratio (an amplifier at the output of a usual interferometer would amplify in the same way signal and noise).

Indeed, if one measures the quadrature variable \cite{c1}, the result is
\be
\langle \Delta^2 X  \rangle= |G_1 G_2 + (G_1^2-1) (G_2^2-1) e^{i (\theta_1 + \theta_2)}|^2 \,\,
\ee
reaching a minimum at the dark fringe $(\theta_1 + \theta_2)= \pi$ for all quadrature variables (at variance with squeezed state interferometry, where only one quadrature experiences noise reduction).

One can then evaluate the relative signal to noise ratio (SNR) for a quadrature measurement for a small change $\delta$ in the phase in one arm is (in the case $(G_2^2-1) >> (G_1^2-1)$) \cite{Li}
\be
SNR = {\langle Y \rangle^2 \over \langle \Delta^2 Y \rangle} = 2 [ G_1 + \sqrt{(G_1^2-1)} ] \cdot |\alpha |^2 \delta^2
\label{SNR}
\ee
which provides a  $[ G_1 + \sqrt{(G_1^2-1)} ]/2$ enhancement respect to a traditional interferometer. A result that is substantially unaffect by detection inefficiencies (or other losses outside the interferometer)  \cite{fwm2,b,l}. On the other hand, it is not insensitive to losses inside the interferometer \cite{c1}.

The first experimental realization of a SU(1,1) interferometer was achieved in Ref.\cite{fwm1}, where the two parametric amplifiers were based on a four-wave mixing process in $^{85}$Rb vapor cells 12 mm long. They were respectively illuminated by an intense vertically polarized pump beam (400 mW)  from a frequency-stabilized Ti:Sapphire laser locked to a Fabry-Perot reference cavity. Furthermore, a  horizontally polarized seed beam was combined with the pump beam at an angle of 0.7° in the center of the first vapor cell. The pump laser frequency was about 0.8 GHz blue detuned from the $^{85}$Rb $\, F=2 \rightarrow F ' $ transition at 795 nm and the seed signal beam was about 3.04 GHz red shifted from the pump with an acoustic optic modulator. The  amplified signal and the conjugate idler beams of the first cell were addressed  into the second vapor cell where they were symmetrically crossed with the pump , similar to the first vapor cell. The two output beams from the second vapor cell were then measured demonstrating a quadratic fringe intensity dependence on the intensity of the phase sensing field at high gain.

\begin{figure}
\includegraphics[height=6cm]{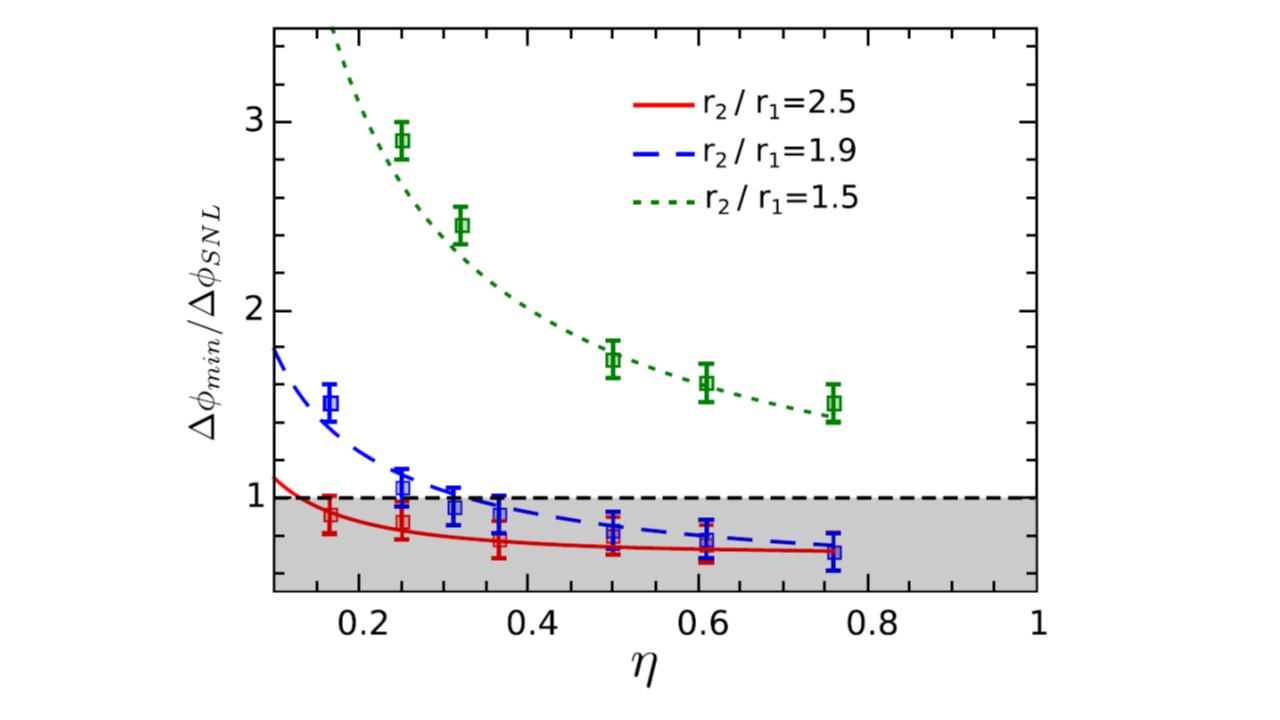}% Here is how to import EPS art
\caption{\label{masha} Reproduced on permission from Ref. \cite{b}: Best phase sensitivity normalized to shot noise
against the detection transmission $\eta$. The curves correspond to different parametric gains. %Red: $r_1 = 2.1, |r_2|= 5.2$. Blue: $r_1 = 2.5, |r_2|= 4.7$. Green: $r_1 = 2.6, |r_2|=4
}
\end{figure}

As a second example of experimental implementation, I will present in some detail the one of Ref.\cite{b}. Here the interferometer consisted of two cascaded 3 mm long $\beta$-barium borate (BBO) crystals (the first placed on a translational stage for controlling the interferometer phase) cut for collinear frequency degenerate type-I phase matching. The pump was the second harmonic of a 400 nm laser with a 5 kHz repetition rate, 1.5 ps pulses. The multimode parametric down-converted (PDC) light generated by the SU(1,1) interferometer was then spatially and spectrally filtered before being detected. In this realisation of the SU(1,1) interferometer both the outputs of the first PA cross the same medium producing the same phase change. An advantage of this configuration is that, since the three fields (idler, signal and pump) copropagate, random phase fluctuations (e.g. caused by air flow, temperature fluctuations,...) wipe out. One of the main results of this work is the demonstration of the insensitivity to losses: for specific gain choices the phase estimation was under shot noise up to 80\% losses, see Fig. \ref{masha}.
This kind of interferometer was then further investigated \cite{masha3,masha4} and, in particular, extended to a wide field SU(1,1) interferometer \cite{masha2}.

The idea of a dual beam interferometer (collinear, but non degenerate PDC with two independent measurements) was then considered theoretically in Ref.\cite{v2} and experimentally in Ref. \cite{v2exp}. Since the SU(1,1) interferometer fringes depend on the sum of the phases acquired by the two beams, this configuration allows optimising the measurement achieving the same SNR of squeezed state interferometry.

In order to point toward real applications of SU(1,1) interferometers, to increase intensity is mandatory. In Ref. \cite{pumpup} a pumped up version of SU(1,1) interferometer using all particles (suited for atomic interferometers) was proposed,  while in Ref.\cite{Liu} a bright seeded SU(1,1) interferometer, where the pump beams generated by a Titanium sapphire laser were injected in $^{85}$Ru vapor cells. This last experiment led to a 1.15 dB advantage respect to standard noise limit, where losses lowered the expected advantage, $\sqrt{G^2(1-G)^2/(G^2+(1-G)^2)}$ (with $1/\sqrt{N}$ scaling),  of 2.6 dB.

An alternative for overcoming the limited intensity power, that remains the most severe drawback of this kind of interferometers, was recently suggested and realized considering a configuration where a SU(2) interferometer is nested in a SU(1,1) one \cite{Du2}. In more detail, the signal beam of the first PA is fed into the dark port of a SU(2) (a Mach-Zehnder) interferometer, while an intense coherent state is fed on the other port. The light emerging from the dark port of this SU(2) interferometer is then recombined with the idler beam in the second PA. The experiment, with two $^{85}$Rb cells generating four wave mixing, led to a 2.2(5) dB advantage, that was substantially maintained  even with a 1mW power.

Concluding this section it is worth mentioning a few further proposals suggesting "variations" on the theme of the SU(1,1) interferometer \cite{v1,Wu,t1,kz,Pau,cav20}.

 In Ref.\cite{v1} the second PA is substituted by a beam splitter (in this case of course idler and signal beam must have the same wave  length) followed by a homodyne detection leading substantially to the same SNR of Eq.\ref{SNR},

 In Ref.\cite{Wu} it was theoretically studied the phase sensitivity of a SU(1,1) interferometer with a coherent state in one input port and a squeezed-vacuum state in the other input
port using the method of homodyne detection, showing the potentiality of achieving the Heisenberg limit.

Finally, in Ref. \cite{t1} it was demonstrated, theoretically and experimentally, that the second PA can eventually be substituted by two homodyne detections (with the same local oscillator) combining the relative photocurrents, while in Ref.\cite{kz} application to stochastic phase estimation was  studied, in Ref.\cite{Pau} a theoretical analysis of a multimode integrated SU(1,1) interferometer was performed,  in Ref.\cite{Liu} the effect of additional external resources was investigated and in the very recent Ref. \cite{2021} it was shown the possibility of combining the two photodetector outputs with an optimal weight factor for beating the shot noise by the same amount regardless of the phase shift in the interferometer.

\section{\label{sec:level1}Quantum enhanced correlation interferometers}

One more significant application of quantum light concerns the possibility of enhancing the performances of correlation interferometry.

 In particular, in the last years the interest for correlation interferometry grew in connection to the possible application to the search of quantum gravity effects. Indeed, the dream of building a theory unifying general relativity and quantum mechanics, the so called quantum gravity (QG), has been a key element in theoretical physics research for the last 60 years:  several attempts in this sense have been considered. However, for many years, no testable prediction emerged from these studies, leading to the common wisdom that this kind of research was more properly a part of mathematics than of physics, being by construction unable to produce experimentally testable predictions as required by Galilean scientific method.
In the last few years this common wisdom was challenged \cite{am1,am2,am3,hog,alt,bek:12}. In particular, it has been proposed that Planck scale effects, connected to non-commutativity of position variables  in different directions \cite{ac,ac2}, could be measured, for instance  in cavities with microresonators \cite{alt} or in two coupled interferometers \cite{hog}, the so called ``holometer''.

This possibility led to the building of the 40m Fermilab holometer~\cite{Hogan2012,Chou2016,Chou2017}, that recently started to pose limits ($2.1 \cdot 10^{-20} m/\sqrt{Hz}$) on the possible noise stemming from these effects \cite{Chou2016,Chou2017} in 1-6 MHz region.

 In synthesis, a holometer is a device consisting of two Michelson interferometers (MI), that are spatially close. The purpose of the holometer is to search for a particular type of correlated noise, which is conjectured to arise from gravitational effects at the Planck scale,  affecting the two spatially close interferometers as relative phase noise.
 A little more in detail, the idea at the basis of these studies is that non-commutativity at the Planck scale ($l_p = 1.616 \times 10^{-35}$ m) of position variables in different directions, predicted by several Planck scale physics models \cite{PL,PL1},  generates an additional very weak phase noise, referred to as holographic noise (HN). In a single interferometer this noise substantially confounds with other sources of noise,
even when the most sensible gravitational wave interferometers are considered \cite{hog}, since by construction their HN resolution is worse than their resolution to gravitational-wave at low frequencies. Nevertheless, when the two equal interferometers  of the holometer have overlapping space--time volumes, then the HN between them is correlated and can be identified easier \cite{hog}. Since the ultimate limit for holometer sensibility, as for any classical-light based apparatus, is dictated by the shot noise, the possibility of going beyond this limit by exploiting quantum optical states is of the utmost interest \cite{giov:11,sch:10,bri:10}.

In addition to this application, measuring relative phase noise could find application in the detection of the gravitational wave background~\cite{Romano2017, Akutsu2008, Nishizawa2008, Shoda2014}, or finding traces of primordial black holes~\cite{Carr1974, Chou2017a}.

The use of quantum light in correlation interferometry was firstly investigated in \cite{Ruo-Berchera2013, Ruo-Berchera2015}.

For this purpose it was considered that the observable measured at the output of the holometer is described by an appropriate operator $\widehat{C}(\phi_1, \phi_2)$, $\phi_k$ being the phase shift (PS) detected by the interferometer $I_k$, $k=1,2$, with expectation value $\langle\widehat{C}(\phi_{1}, \phi_{2}) \rangle=\mathrm{Tr}[\rho_{12}\widehat{C}(\phi_{1}, \phi_{2}) ]$, where
$\rho_{12}$ is the overall density matrix associated with the state of the two light beams injected in $I_1$ and $I_2$.
\par
In the Fermilab scheme the idea for observing the eventual existence of the HN is comparing $\langle\widehat{C}(\phi_{1}, \phi_{2}) \rangle$ in two different experimental configurations of $I_1$ and $I_2$, namely when the arms are parallel, ``$\parallel$'', or perpendicular, ``$\perp$'' \cite{hog}. Indeed, according to Ref.\cite{hog}, one expects that HN is correlated in the first case (being the arms in the same light cone), while it is not in the second case. Thus, on the one hand, in the configuration``$\parallel$'' the correlation of the interference fringes would highlight the presence of the HN. On the other hand, the configuration ``$\perp$'' serves as a reference measurement, namely, it corresponds to the situation where the correlation due to HN is absent, in other words, it is equivalent to estimating the ``background''.

In general, being the statistical properties of the PSs fluctuations due to HN described by a suitable probability density function, $f_x (\phi_{1}, \phi_{2})$, $x=\parallel,\perp$, the expectation of any operator $\widehat{O}(\phi_1,\phi_2)$ is obtained by averaging over $f_x$, namely, $\langle\widehat{O}(\phi_1,\phi_2)\rangle \to
\mathcal{E}_{x}\left[\widehat{O}(\phi_1,\phi_2)\right]
\equiv \int \langle\widehat{O}(\phi_1,\phi_2)\rangle ~ f_x (\phi_{1}, \phi_{2})  ~
\mathrm{d} \phi_{1} ~ \mathrm{d} \phi_{2}$.
\par
Since in the holometer the HN can be observed through a correlation between two phases, the appropriate function to be estimated is their covariance:
\begin{equation}\label{cov:PS}
\mathcal{E}_\parallel\left [ \delta \phi_{1} \delta \phi_{2}\right] \approx
\frac{
\mathcal{E}_\parallel \left[ \widehat{C}(\phi_1,\phi_2)\right]-
\mathcal{E}_\perp \left[ \widehat{C}(\phi_1,\phi_2)\right]}
{\langle \partial_{\phi_{1},\phi_{2}}^{2}
\widehat{C}(\phi_{1,0}, \phi_{2,0}) \rangle }, \quad
(\delta \phi_{1},\delta\phi_{2} \ll 1)
\end{equation}
$\delta \phi_{k} = \phi_k - \phi_{k,0}$, $\phi_{k,0}$ being the mean PS value measured by $I_k$, $k=1,2$ reducing as much as possible the associated uncertainty:
\begin{equation} \label{U1} \mathcal{U}^{(0)} = \frac{\sqrt{2\, \mathrm{Var}\left[ \widehat{C}(\phi_{1,0 }, \phi_{2,0}) \right]}}{\left|
 \langle \partial_{\phi_{1},\phi_{2}}^{2} \widehat{C}(\phi_{1,0},
 \phi_{2,0}) \rangle \right|},
\end{equation}
where $\mathrm{Var}\left[\widehat{C}(\phi_{1,0 }, \phi_{2,0}) \right]=
\langle
\widehat{C}(\phi_{1,0}, \phi_{2,0})^2 \rangle - \langle
\widehat{C}(\phi_{1,0 }, \phi_{2,0}) \rangle^2 $ does not depend on the
PSs fluctuations due to the HN, but it represents the intrinsic
quantum fluctuations of the measurement described by the operator
$\widehat{C}(\phi_1,\phi_2)$ and depends on the optical (quantum) states
entering the holometer.

In \cite{Ruo-Berchera2013, Ruo-Berchera2015} it was demonstrated that, when the two input modes of each interferometer $I_k$, $k=1,2$, are excited in a coherent state and a squeezed vacuum state with mean number of photons $\mu$ and $\lambda$, respectively:
\begin{equation}\label{U1sq}
\mathcal{U}_{\rm SQ}^{(0)} (\mu,\lambda)\approx\sqrt{2}~\frac{\lambda + \mu \left(1+2 \lambda - 2 \sqrt{\lambda
+\lambda^2}\right)}{(\lambda-\mu)^2}.
\end{equation}
As expected, in analogy with the PS measurement for a single interferometer \cite{Int}, if $\mu \gg \lambda \gg 1$, then we have the optimal accuracy $\mathcal{U}_{\rm SQ}^{(0)} \approx (2\sqrt{2} \lambda \mu )^{-1}$. This represents an evident advantage in terms of uncertainty reduction (of the order $(4 \lambda )^{-1}$) with respect to classical case $\mathcal{U}_{\rm CL}^{(0)}\approx \sqrt{2}/ \mu $ when only coherent states are employed.
An important difference arises between the single interferometer PS measurement, involving a first order moment of the photon number distribution,  and the covariance estimation, involving the second order moments: whilst in the first case the uncertainty scales as the usual standard quantum limit one, $\propto\mu^{-1/2}$, in the second case it scales  $\propto\mu^{-1}$ (neglecting the little relative contribution of the squeezed state to the intensity).

Nonetheless, even a larger advantage was demonstrated when considering a  configuration  where second port input modes of $I_1$ and $I_2$, $a_1$ and $a_2$, are the component of a photon-number entangled beam, in particular a twin beam (or a two-mode squeezed vacuum state) \cite{rev2}
\be
|{\rm TWB}\rangle\rangle = (1+\lambda)^{-1/2} \sum_{n}(1+\lambda^{-1})^{-n/2} |n\rangle_{a_1} |n\rangle_{a_2}
\label{twb}
\ee
 while two coherent states still enter into the other two ports.  This correlation property leads to the amazing result that the contribution to the uncertainty coming from the photon number fluctuation noise  is $\mathcal{U}_{\rm TWB}^{(0)}=0$ (when $\lambda,\mu\ne0$) in the ideal situation of no losses, representing an ideal accuracy of the interferometric scheme to the PSs covariance due to HN.
\par

This advantage is conserved even in presence of losses in the case of high quantum resources exploited, {\it i.e.} $\mu\gg\lambda\gg1$, where one finds
$\mathcal{U}_{\rm SQ}^{(0)}/\mathcal{U}_{\rm CL}^{(0)} \approx \left( 1-\eta\right)+\eta/(4\lambda)$ and
$\mathcal{U}_{\rm TWB}^{(0)}/\mathcal{U}_{\rm CL}^{(0)} \approx 2\sqrt{5}\left( 1-\eta\right)$
when the total detection efficiency $\eta$ is sufficiently large.

These theoretical ideas were then experimentally demonstrated in \cite{holexp}.

Both the two independent squeezed state and a twin-beam like injection were considered. Each Michelson interferometer (MI), with arm length of 0.92m, was fed with 1.5 mW of 1064 nm light from a low noise Nd:YAG laser source. This laser source was also used to seed the optical parametric oscillators generating the squeezing. The light for the MIs was spatially cleaned with an optical fiber. Then, each interferometer had two piezo-actuated high-reflectivity end mirrors ($R_\textrm{M} = 99.9 \ \%$), while the partially reflecting mirrors for the two power recycling cavities had a reflectivity of $R_\textrm{PRM} = 90 \ \%$.

The squeezed-light (6.5 dB relative to the shot noise), produced by a parametric down-conversion in a potassium titanyl phosphate (PPKTP) crystal placed in a semi-monolithic linear cavity, was injected in the two MIs via their output ports.

 The signal-to-noise ratio  with squeezing injected was consistently higher, by a factor of 2, of the classical interferometer.
In the spectral domain, correlated signals were extracted by the cross-linear spectral density of the two interferometers, demonstrating an improvement of a factor 1.35  from the injection of squeezed states. A result that also poses an upper limit on HN in the 13 MHz region ($3 \cdot 10^{-17} m/\sqrt{Hz}$).

\begin{figure}
\includegraphics[height=6.2cm]{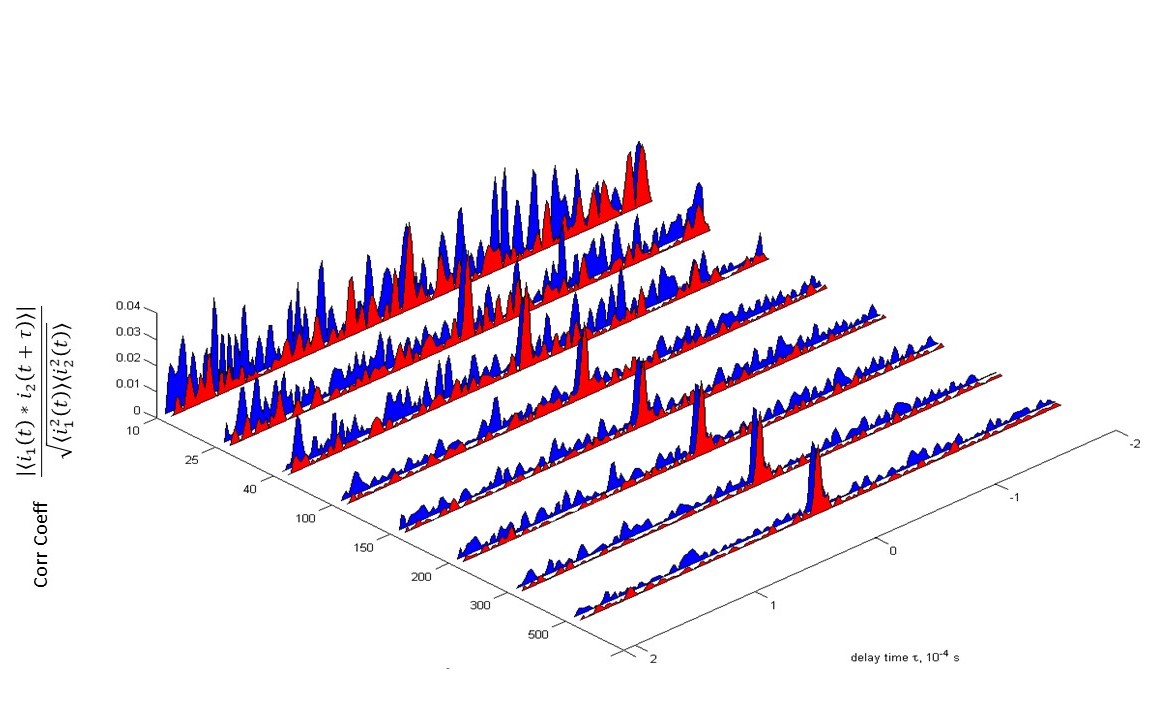}% Here is how to import EPS art
\caption{\label{psd} Analysis of the time series obtained from the independent squeezer configuration from the experiment of Ref.\cite{holexp}, with added white noise. The red curves correspond to the squeezing injection configuration and the blue curves to the coherent case.  The plot shows the cross-correlation of the data in the time domain versus the number of samples. The correlation peak is hidden in the noise for small sample numbers, equivalent to short acquisition time, and it emerges when the number of samples is increased. The use of squeezing allows clearly for an earlier detection of the peak.}
\end{figure}

\begin{figure}
\includegraphics[height=7cm]{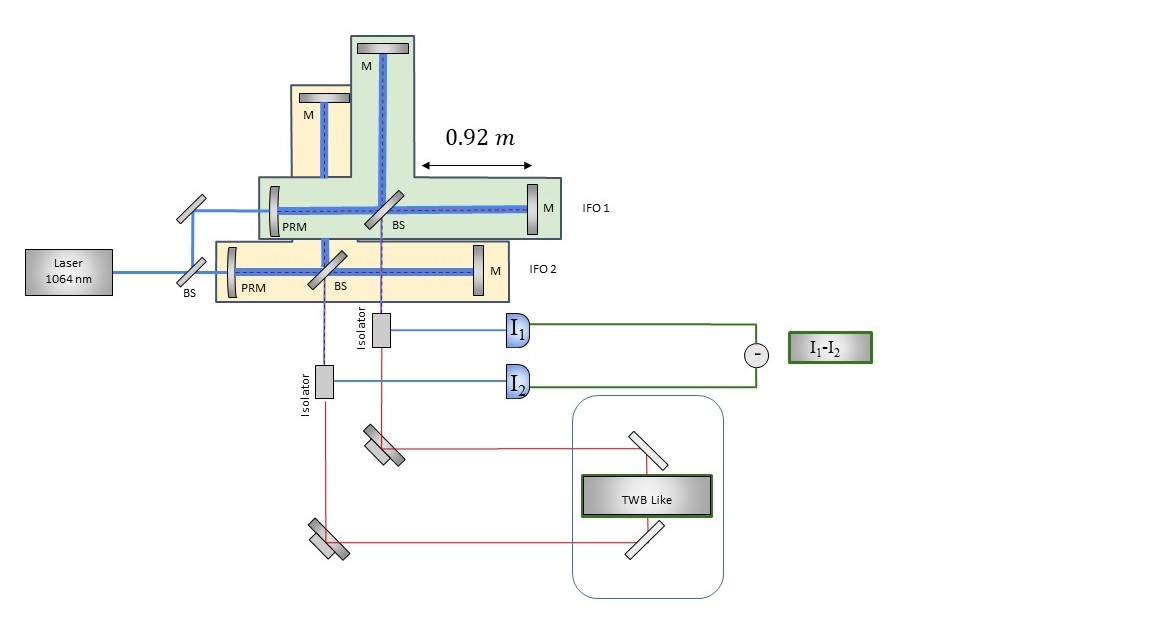}% Here is how to import EPS art
\caption{\label{setup} Simplified schematic of the double-interferometer setup of Ref\cite{holexp} in the twin beams configuration. The two spatially close ($\sim$ 10 cm) interferometer input ports were fed 1064 nm laser light from a low noise Nd:YAG laser source. Each interferometer had a power recycling mirror (PRM) in the input port, to form a cavity around the interferometer. The input beams were split at a beamsplitter (BS) in each interferometer, and subsequently impinged on piezo controlled end mirrors (M). A Faraday isolator in each output port allowed for measuring the output while  twin-beam (TWB) modes were injected into the output ports. }
\end{figure}

When a twin-beam like signal was injected (see Fig.\ref{setup}),  the correlation between the two modes led to a noise reduction of 2.5 dB with respect to the SNL.
\begin{figure}
\includegraphics[height=6cm]{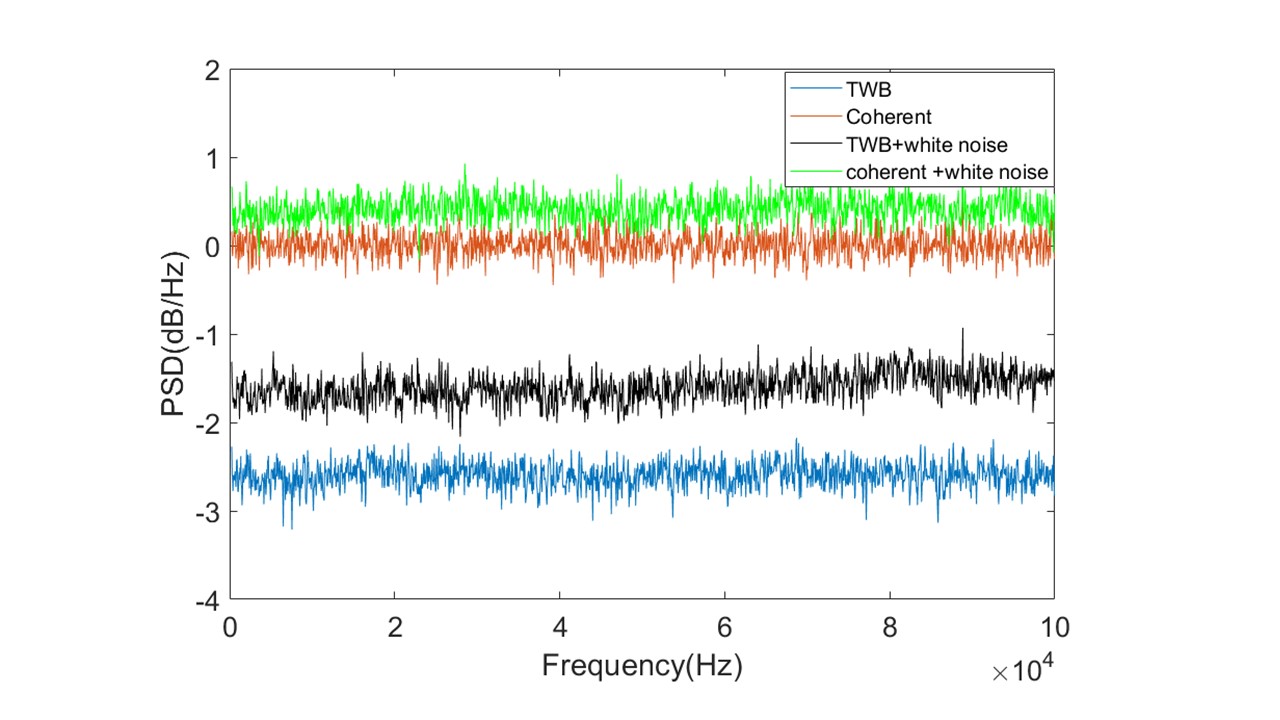}% Here is how to import EPS art
\caption{\label{psd} Comparison of the PSDs of the photocurrent difference between the MI outputs of the experiment of Ref.\cite{holexp}, with and without TWB. Green and black traces have uncorrelated white noise introduced in the MIs, with and without TWB respectively. The white noise clearly emerges in the TWB case demonstrating the efficiency of this method.}
\end{figure}

This enhancement was also observed in the power spectral density(PSD), fig.\ref{psd}, of the subtracted interferometer outputs, demonstrating that the presence of faint uncorrelated noise can be more easily detected by twin beam-like correlations and, therefore, the possible advantage offered by this technique in experiments of correlation interferometry.

Following this proof of principle experiment, very recently in Ref.\cite{holnew} a detailed description  was presented of a table top quantum enhanced holometer (in the version based on the injection of two independent squeezed beams), with the purpose of realising a new experiment on Planck scale effects search.

 Furthermore, interest emerged in investigating possible advantages of using non-gaussian (as photon subtracted) states \cite{nig}, while in Ref.\cite{Ben} decoherence effects in this kind of experiments were analyzed.

\section{\label{sec:level1}New ideas and perspectives}

Finally, in this section we consider new schemes and ideas that still far from practical applications, could eventually lead to interesting developments in  a more or less next future.

A first line of research, somehow connected to SU(1,1) interferometers,  is inserting PA in the arms of an interferometer for enhancing its performances. A first example was experimentally realised in Ref.\cite{dema}, where the performances of a Mach-Zehnder interferometer operating with single photons probes were enhanced by inserting a parametric amplifier in one arm, still scaling as $1/\sqrt{N}$. Very recently a new experiment with two PAs, one per arm, was realised \cite{zuo}, achieving a 5.6 dB squeezed noise floor under shot noise limit and a 4.9 dB enhancement of signal to noise ratio, reaching Heisenberg limit.

Another very interesting seminal proposal \cite{NOON1,NOON2}, much discussed in the theoretical literature, is the use of photon number entangled states of the form
\be { |N \rangle |0 \rangle + |0
\rangle |N \rangle \over \sqrt{2} }
\ee
dubbed NOON states.

In synthesis, while using N independent photons in a two arms (a,b) interferometer one has
\be
[{ |a \rangle  + e^{i  \theta} |b \rangle \over \sqrt{2} } ]^{\otimes N}
\ee
leading to a as $1/\sqrt{N}$
  phase uncertainty scaling,   with a NOON state the phase cumulates as
\be
{ |N \rangle_a |0 \rangle_b + e^{i N \theta} |0
\rangle_a |N \rangle_b \over \sqrt{2}}
\ee
leading to a $1 + \cos(N \theta)$ interference dependence, which implies a $1/N$ scaling \cite{Huelga}.
Albeit of large theoretical interest, unluckily no effective idea exists for creating high N photon NOON states with present technology, and only a few proof of principle experiments were realised with N=2 \cite{N1,N2,slu,N3,NN1,NN2} (easily created by Hong-Ou-Mandel interference) or N=4,5 \cite{N4,N5} (with postselection that, due to low probability success, limits an eventual practical use). Furthermore, the advantage offered by these states is very sensitive to losses \cite{kn}.

Also the use of twin photon Fock states $| N,N \rangle$, would lead to asymptotically approaching Heisenberg limit \cite{HB,mig,ok}. Nonetheless also high N photon Fock states are far to be produced \cite{rad} (postselected heralded N=8 Fock states are the present limit \cite{n8}), as well as maximally correlated states defined in Ref. \cite{Bo}, which would allow exactly reaching Heisenberg limit.

In order to overpass these difficulties, on the one hand the use of twin beams coupled to photo-count measurements was suggested \cite{du} and, in a simplified version,  implemented \cite{mat}. In this proof of principle experimental realisation the photon counting detection was realised through a multiplexed scheme; the measured data were then compared with what expected by a different phase shift impressed on the  two components of the twin beam and a final detection  described by a specific POVM keeping into account the multiplexed measurement. The shot noise limit was surpassed despite a realistic level of losses. Very recently \cite{ger} a new experiment enhanced further the advantage by using NIST transition edge sensors.

 On the other hand new detection methods, suited for reaching Heisenberg limit with quantum states experimentally realisable, were considered. Heisenberg limit in phase estimation was also reached in Mach-Zehnder interferometers  by exploiting a generalized Kitaev's algorithm, reaching a 10dB advantage respect to standard quantum limit with highest number of resources (378 photons) \cite{Pry}. Nevertheless, the extension to practical uses (i.e. a much larger number of photons) still remains  far from reach.

A further class of proposals is based on using parity measurements \cite{par}, i.e. on measuring the parity operator $\Pi = e^{i \pi a^\dag a}$, which substantially requires photocounting detection. A first proposal in this sense appeared in Ref.\cite{Dow1}: here it  is suggested to use collinear degenerate twin beams, Eq. \ref{twb}, i.e.two mode squeezed vacuum (TMSV),  entering one port of the interferometer, e.g. 1 in Fig.\ref{mz}, and performing a parity measurement to one output port, e.g. 4 in Fig.\ref{mz}. This leads to:
\be
\langle \Pi \rangle_{\theta + \pi/2} = {1 \over \sqrt{1+\bar{n}(\bar{n}+2) \sin^2\theta }}
\ee
where $\bar{n}$ is the average photon number of the TMSV, which, through Cram\'er - Rao bound, leads to a $1 / \sqrt{ \bar{n} (\bar{n}+2)}$ dependence for the phase uncertainty in proximity of $\theta = 0$, that even beats the Heisenberg limit: here is an example of application of Hofmann limit \cite{Hof}, since  $\triangle n^2 = \langle n^2 \rangle -\langle n \rangle^2 = \bar{n}^2 + 2 \bar{n}$ for TMSV.

Heisenberg limit can eventually be beaten also through non-linear transformations \cite{Luis}, i.e. this limit can be overcome when the generator of the transformation is proportional to the photon number, e.g. in \cite{Luis2} a Kerr type $e^{i \theta (a^\dag a)^2}$ transformation was considered leading to a $\triangle \theta \sim 1/N^{3/2}$  scaling, besides a finite detection efficiency only represents a multiplicative factor to this scaling. More in general a Hamiltonian including all possible k-body couplings allows a $1/N^k$ scaling for entangled probe states,  $1/N^{k-1/2}$ for a probe initially in a product state \cite{Luis,Luis2,Bo1,Bo2}, representing in this last case also an interesting example of quantum enhanced measurement without entanglement \cite{P,B}. Further theoretical optical schemes with non linear phase shifts can be found in Ref. \cite{Sp,Yu}.

Finally, different ideas emerged on using weak values  or weak values-inspired methods \cite{2,p} in interferometry.
Weak value measurements were introduced \cite{2} in 1988 Aharonov, Albert and Vaidman, and   they represent a new paradigm of quantum measurement where so little information is extracted from a single measurement that the state does not collapse.
In little more detail, the weak value of an observable $\widehat{A}$ is defined as $ \langle \widehat{A} \rangle _w = { \langle \psi_f | \widehat{A} |\psi_i \rangle \over \langle \psi_f | \psi_i \rangle}$, where the key role is symmetrically played by the pre-selected ($ | \psi_i \rangle$) and post-selected ($|\psi_f \rangle$) quantum states.
When considering a von Neumann coupling between the observable $\widehat{A}$ and a pointer observable $\widehat{P}$, according to the unitary transformation $\widehat{U}= \exp (- i g \widehat{A} \otimes \widehat{P})$, in the weak interaction regime the evolution of this system is
\begin{equation}
\langle \psi_f | e^{- i g \widehat{A} \otimes \widehat{P}}  |\psi_i \rangle \simeq  \langle \psi_f  |\psi_i \rangle (\mathbf{1} -i g \langle \widehat{A} \rangle_w \widehat{P}) .
\end{equation}
Thus, weak values are measurable quantities and can provide an "amplification" in measuring a small parameter g (e.g. a phase) \cite{2,3,4,5,6,7,8,9,10,11}, which can be significantly useful in presence of technical noise \cite{boyd}.

For example, in Ref.\cite{how} it was demonstrated (both theoretically and experimentally in a Sagnac configuration) that it is possible reaching the same sensitivity of balance homodyne detection measuring the dark port only. This could allow a general improvement of sensitivity since allowing a large reduction of intensity at the detector. This method was recently extended in Ref. \cite{zh}.

As mentioned, weak value amplification finds a significant use in presence of technical noise. On this line, in Ref.\cite{Li2} it was demonstrated that they can be applied to reduce systematic uncertainties in phase estimation. Specifically, it was considered the case of a Sagnac interferometer where the beam splitter is substituted by a polarizing beam splitter and the phase is encoded in one of the polarizations. The experiment demonstrated a 30 times phase resolution improvement. A further approach exploiting weak values appeared in Ref.\cite{zeng}. Here it is considered a Michelson (or Mach - Zehnder) interferometer with a "stable" polarization dependent phase shift in one arm. In this case the preselected state is (in terms of Horizontal, H, and vertical, V, polarization states) of the form ${(| H \rangle  + | V \rangle) \over \sqrt{2}}$ and the postselection is $(cos \alpha| H \rangle  + sin\alpha | V \rangle) $, while the polarization measurement operator is, for instance, $\widehat{A} = |H\rangle \langle H|$. The uncertainty is amplified as the signal by the weak value $\langle \widehat{A} \rangle_w$, but a small weak value attenuates phase and uncertainty at the same time. Thus, the reduction of the uncertainty is achieved by exploiting a feedback control: a $\sim 10^{-3}$ reduction factor was experimentally demonstrated.

Thus, in summary, several new ideas are emerging for further improving optical (and not only) interferometry. Many of them already were demonstrated in proof of principle experiments. However, if and which of them will really find practical applications, overcoming present limitations, still remains to be clarified.

\section{\label{sec:level1}Conclusions}

In the panorama of quantum technologies, quantum metrology represents one of the most significant candidates to practical applications in an advanced stage. In particular quantum enhanced optical interferometry has already found significant use in gravitational wave detectors.  Furthermore, several other methods have already overpassed the stage of theoretical proposals having been realised at least in proof of principle experiments, that are prompting new applications of these methods, for example in correlation interferometry.

The theoretical and experimental studies in this area are further promoted by new proposal of applications that range from technological needs, in sensing or computation or other, to fundamental physics, some already significant nowadays, as gravitational wave detection, other addressing more visionary ideas that could be of interest in a next or farther future, ranging from Planck scale effects \cite{Hogan2012,mar,g1,g2} up to wormhole search \cite{worm}, ...

\begin{acknowledgments}
This work has received funding from the European Commision’s PATHOS EU H2020 FET-OPEN grant no. 828946 and
Horizon 2020, from the EMPIR Participating States in the context of the project 17FUN01 “BeCOMe”.
\end{acknowledgments}

\section{Data availability statement}

Data sharing not applicable – no new
data generated
\appendix
\vskip 0.5cm
The following article has been submitted  by AVS Quantum Science. After it is published, it will be found at  https://avs.scitation.org/journal/aqs.


\begin{thebibliography}{1}
\bibitem{losses} Incidentally losses are usually significantly affecting all quantum technologies protocols (with the significant exception of quantum illumination \cite{qillu,qillu1,qillu2}).
\end{thebibliography}
\end{document}